\documentclass[aps,pre,twocolumn,superscriptaddress,showpacs]{revtex4-1}
\usepackage{graphicx}
\usepackage{amsmath}
\usepackage{natbib}
\usepackage[usenames]{color}
\usepackage{soul}

\usepackage{amsfonts}
\usepackage{dcolumn}                    	     
\usepackage{amssymb}

\usepackage{bm,fancybox}                	     

\begin{document}
\title{Classical transients and the support of open quantum maps}

\author{Gabriel G. Carlo}
\email[E--mail address: ]{carlo@tandar.cnea.gov.ar}
\affiliation{Departamento de F\'{\i}sica,
 Comisi\'on Nacional de Energ\'{\i}a At\'omica.
 Av.~del Libertador 8250, 1429 Buenos Aires, Argentina}

\author{D. A. Wisniacki}
\affiliation{ Departamento de F\'\i sica and IFIBA,
FCEyN, UBA Ciudad Universitaria,
Pabell\'on 1, Ciudad Universitaria, 1428 Buenos Aires, Argentina}

\author{Leonardo Ermann}
\affiliation{Departamento de F\'{\i}sica,
 Comisi\'on Nacional de Energ\'{\i}a At\'omica.
 Av.~del Libertador 8250, 1429 Buenos Aires, Argentina}

\author{R. M. Benito}
\affiliation{Grupo de Sistemas Complejos
 and Departamento de F\'{\i}sica,
 Escuela T\'ecnica Superior de Ingenieros
 Agr\'onomos, Universidad Polit\'ecnica de Madrid,
 28040 Madrid, Spain}

\author{F. Borondo}
\affiliation{Departamento de Qu\'{\i}mica, and
 Instituto de Ciencias Matem\'aticas (ICMAT),
 Universidad Aut\'onoma de Madrid,
 Cantoblanco, 28049--Madrid, Spain}
\date{\today}
\begin{abstract}
The basic ingredients in a semiclassical theory are the classical invariant 
objects serving as a support for the quantization.
Recent studies, mainly obtained on quantum maps, have led 
to the commonly accepted belief that it is the classical repeller -- 
the set of non escaping orbits in the future and past evolution -- 
the object that suitably plays this role in open scattering systems.
In this paper we present numerical evidence warning that this may not always 
be the case. 
For this purpose we study recently introduced families of 
tribaker maps [L. Ermann, G.G. Carlo, J.M. Pedrosa, and M. Saraceno, 
Phys. Rev. E {\bf 85}, 066204 (2012)], which share the same asymptotic properties but 
differ in their short time behavior. 
We have found that although the eigenvalue distribution 
of the evolution operator of these maps follows the fractal Weyl law 
prediction, the theory of short periodic 
orbits for open maps fails to describe the resonance eigenfunctions of some of them. 
This is a strong indication that new elements must be included in the semiclassical 
description of open quantum systems.
\end{abstract}
\maketitle

%
\section{Introduction}
 \label{sec:intro}

The strong recent interest in the study of open quantum systems has
evidenced that this area is still in a very early stage of development,
specially when compared with the knowledge that we have of the closed 
counterparts.
As a consequence, it 
has become an extremely active topic in fundamental physics \cite{Weiss}.
The most important advance
concerns scattering systems, 
where the fractal Weyl law (FWL) for the number of long-lived resonances has
been conjectured and tested 
in different examples \cite{Nonnenmacher}. 
The corresponding classical invariant distribution, 
i.e. the fractal hyperbolic set of all the trajectories
non-escaping in the past and future (the repeller), plays a
fundamental role 
for the interpretation of the quantum spectrum of
quasibound (resonant) states. 
According to this law \cite{conjecture} the number of long-lived resonances 
scales as $N^{d/2}$, where $d$ is the fractal dimension
of the repeller and $N$ that of the Hilbert space. 
After the proof of the fractal Weyl upper bound for a Hamiltonian flow
showing a fractal trapped set \cite{Sjostrand}, 
results that support the validity of the conjectured law
both for smooth and hard 
wall potentials \cite{Hamiltonians} have followed. 
However, open quantum maps provide the ideal arena 
for this kind of studies, due to its computational
simplicity \cite{qmaps}, and accordingly we will focus on them in this work. 

Another recent advance in the understanding 
of scattering systems has been the development of the short periodic orbits (POs) 
theory for open quantum maps \cite{art1}. 
This theory makes use of the shortest POs living 
in the repeller to construct a classically motivated basis in which 
the quantum non unitary operators corresponding to the classical maps
can be adequately expressed. 
It has been shown that there is a connection between 
the FWL and the possibility to obtain the quantum long-lived resonances by 
using the detailed classical information of the repeller, 
i.e.~the embedded trajectories \cite{art2}. 

As a result of all this evidence, the repeller is commonly accepted 
to be the fundamental invariant classical structure 
in phase space able to explain the quantum mechanics of open systems.
However, in a recent publication \cite{art3},
it has been shown for a set of maps sharing 
the same repeller and asymptotic features (including the escape rate) 
but having different classical transient behavior, 
that the quantum properties of the resonances can differ strongly from this view. 
This suggests that also transients may play a fundamental role in the 
description of  open quantum systems.

In this paper we further investigate this issue, by presenting a twofold
numerical calculation on the tribaker map families introduced in \cite{art3}. 
These families differ in the way in which
the openings are defined. 
In the so called shift family the area of the opening is the same 
for all of its members, while in the intersection family this area changes. 
First, we have verified that the resonance distributions follow 
the FWL for these two kinds of maps. 
Second, we have tested 
the ability of the theory of short POs
for open quantum maps to reproduce the eigenvalues and the corresponding 
resonance eigenfunctions.  
This approach is used as a tool to investigate the actual 
classical support of the quantum resonances, finding that 
the quantum information of some of the shift family members
cannot be reproduced. 
These results cast doubts on the commonly accepted belief that the 
repeller is just enough to support the quantization of open maps.

The organization of this paper is as follows: 
In Sec. \ref{sec:maps} we introduce the features of the shift and intersection 
families of open tribaker maps. 
In Sec. \ref{sec:statistics} we study the effects of classical transients on the spectral statistics. 
The same is done in Sec. \ref{sec:shPOth} 
 regarding the classical support of open quantum maps 
by using the theory of short POs as the main tool. 
Finally, in Sec. \ref{sec:conclusions} we present a final discussion, 
and our perspectives for future work.

\section{Open maps: shift and intersection families}
 \label{sec:maps}

Open maps can be defined as the deterministic evolution corresponding to a given (closed) map 
followed by the loss of trajectories which pass through a given region in phase-space.
The set of trajectories that survive in the future and in the past define the backwards and 
forward trapped sets, respectively. In turn, the repeller is given by their intersection.
The structure of this classical invariant can be very complex for chaotic dynamics, 
having a fractal dimension in most of the cases. 
Another relevant property of open maps is the decay rate, 
which is defined as the asymptotic loss of probability in one iteration.

In this work we will study the shift and intersection families of open tribaker maps 
defined in \cite{art3} that differ only in their short time dynamics, sharing the same repeller and decay rate.
They are defined preserving parity symmetry, and therefore with the same openings in position and momentum. 

\begin{figure}
  \includegraphics[width=8cm]{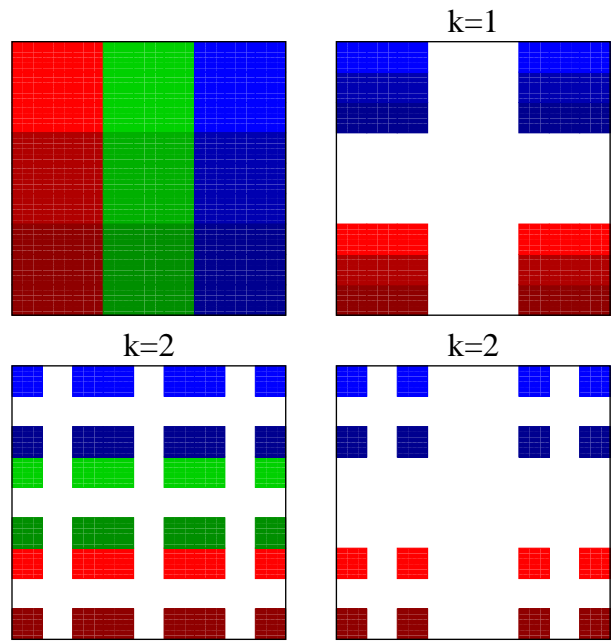}
 \caption{(color online) Iteration of the classical open baker map. The unit square 
 divided in 9 regions (3 colors with different intensities) is shown in the top-left panel.
One iteration of the colored unit square with the open baker map is shown 
for $k=1$ (shift or intersection family) in top-right panel
and for $k=2$ for the shift (bottom-left panel) and intersection (bottom-right panel) cases.}
 \label{fig:0}
\end{figure}

In ternary notation the map action is given by a Bernoulli shift of 
$q=0.\epsilon_0\epsilon_1\epsilon_2\ldots$ and 
$p=0.\epsilon_{-1}\epsilon_{-2}\epsilon_{-3}\ldots$ 
(given by the corresponding trits $\epsilon_j=0,1,2$)
as $(p\vert q)=\ldots\epsilon_{-2}\epsilon_{-1}.\epsilon_{0}\epsilon_{1}\epsilon_{2}\ldots\rightarrow$
$(p^\prime\vert q^\prime)=\ldots\epsilon_{-2}\epsilon_{-1}\epsilon_{0}.\epsilon_{1}\epsilon_{2}\ldots$
where the dot is moved one position to the right. 
Shift and intersection families of open maps can be straightforwardly defined 
replacing the usual trit $\epsilon_j$ by the open trit  $\tilde{\epsilon}_j$ 
where the value 1 is forbidden (i.e., $\tilde{\epsilon}_j=0,2$).
In this framework, the shift family members ($\mathcal{B}^{s}_k$) are defined with two open trits 
corresponding to the $k$-th most significant trit of both position and momentum.    
Equivalently the intersection family members ($\mathcal{B}^{i}_k$) have the first $k$ trits open, 
both in position and momentum.
The iterations of open baker maps in the unit square for $k=1,2$ and both, shift and intersection 
families are illustrated in Fig.\ref{fig:0}. 

The quantum version of the shift and intersection map families can be 
defined from the usual quantum tribaker map in a Hilbert space spanned by $l$ qutrits (with dimension $N=3^l$)
\begin{equation}\label{eq.baker}
B_{pos}=G^\dagger_N B_{mix}=G^\dagger_N \left(\begin{array}{ccc}
 G_{N/3}&0&0\\0&G_{N/3}&0\\0&0&G_{N/3}\\\end{array}\right)
\end{equation}
where the antiperiodic Fourier transform $G_N$ connects position and momentum eigenvectors 
($\vert q_j\rangle$ and $\vert p_j\rangle$ with $j=1,\ldots,N$)  
\begin{equation}\label{eq.ft}
 \left(G_N\right)_{j^\prime,j}\equiv\langle q_{j^\prime}\vert p_j\rangle=\frac{1}{\sqrt{N}}
e^{-i\frac{2\pi}{N}(j^\prime+\frac{1}{2})(j+\frac{1}{2})}.
\end{equation}
with $N=1/(2\pi\hbar)$ (see \cite{bakerbvs,ermannessential} for quantization details).

The forbidden $1$ in one trit $\tilde{\epsilon}$, can be quantized by means of the one qutrit 
projector $\pi=I-\vert1\rangle\langle 1\vert=\vert0\rangle\langle 0\vert+\vert2\rangle\langle 2\vert$. 
In this way, the projector applied to the $i$-th qutrit can be written as 
\begin{equation}
\Pi_i=\underbrace{I\otimes \ldots \otimes I}_{i-1}\otimes \pi \otimes 
\underbrace{I\otimes \ldots \otimes I}_{l-i}  
\end{equation}
with $\Pi_i=\Pi_i^\dagger$, $\Pi_i^2=\Pi_i$ and $[\Pi_{i},\Pi_{j}]=0$ where $i,j=1,\ldots,l$.
Following this notation, the shift and intersection families of open quantum tribaker 
maps can be straightforwardly defined as
\begin{eqnarray}
 \tilde{B}^{s}_k&=&G^\dagger_N\Pi_{k}B_{mix}\Pi_{k}\\
 \tilde{B}^{i}_k&=&G^\dagger_N
\Pi_{1}^\dagger
\ldots\Pi_{k-1}^\dagger\Pi_{k}^\dagger
B_{mix}\Pi_{k}\Pi_{k-1}\ldots\Pi_{1}
\end{eqnarray}
with $k=1,\ldots,l$, and where the form $\tilde{B}_{mix}=\Pi B_{mix}\Pi$ preserves the parity 
symmetry of the closed map.

\section{Transient effects and the statistics of eigenvalues of open quantum maps}
 \label{sec:statistics}

The geometry of phase space imposes restrictions in the asymptotic distributions of 
the eigenstates of quantum systems. In bounded systems of $f$ degrees of freedom, 
the well known Weyl law establishes that the number of eigenstates $N(E)$ of energy 
less than $E$ is given
by the number of cells of size $h^f$ contained in the corresponding allowed 
volume of phase space $V(E)$, i.e. 
\begin{equation}
N(E)=V(E)/(2 \pi \hbar)^f.
\end{equation}

In open systems the relation between the geometry of the phase space
and the statistics of eigenvalues is more involved. In these systems, the quantum
evolution is given by a nonunitary operator with right and left decaying
nonorthogonal eigenfunctions. 
The corresponding eigenvalues $z_n$ are complex numbers that 
fall inside the unit circle, that is, $|z_n|^2 \le 1$. 
The resonances are conjectured to be localized on the trapped sets 
\cite{ChaoticScattering,ShepelyanskyPhD99, KeatingPrl06}. 
This structure of the phase space has led to the prediction of a relation between 
the number of long-lived resonances and the fractal dimension $d$ of the repeller. 
As previously mentioned, the FWL establishes that the number of resonances with 
decay factor $\mu_n=|z_n|=\exp[-\Gamma_n/2]$ scales as $N_{\mu} \sim N^{d/2}$. 
This conjecture has been investigated in many systems,
including quantum maps, 2-D and 3-D billiards, 
and the modified Henon-Heiles potential \cite{conjecture,qmaps,Non1,Hamiltonians}.
Also, more scarring has been predicted in open systems than in closed ones \cite{Wiersig,carlo}; 
this phenomenon has deep consequences in many applications, 
such as microlasers \cite{Microlasers}.

\begin{figure}
  \includegraphics[width=8cm]{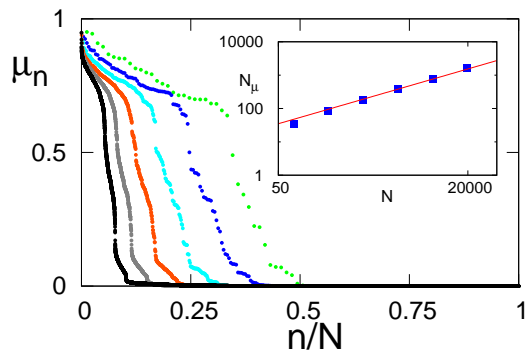}
 \caption{(color online)  Ordered decay factor $\mu_n$ of the usual open
 tribaker map as a function of $n/N$. 
 Hilbert dimensions $N=3^l$ with $l=4,5,6,7,8$ and $9$ (green, blue, turquoise, red, gray, and black symbols, 
respectively). 
 Inset: Fraction of states $N_{\mu}$  with decay rate $\Gamma>0.1$ as a function of the 
Hilbert space dimension. The solid
red line corresponds to the prediction of the FWL ($N_\mu \sim
N^{\ln{(2)}/ (2 \ln{(3)})}$).}
 \label{fig:1}
\end{figure}

The goal of this Section is to test the validity of the FWL
in the families of open tribaker maps introduced earlier. We remark that all these systems
have the same repeller, classical information that determines the statistical 
behavior of the resonances which is embodied in the FWL.
In Fig. \ref{fig:1} we show the ordered decay factor $\mu_n$ 
for the usual open tribaker map 
(i.e., the first member of any of the two families, $\tilde{B}_1^{s/i}$), 
having Hilbert space dimensions $N=3^l$ with $l=4,5,6,7,8$ and $9$. In the inset of Fig.
\ref{fig:1} we show the number of resonances
with $\Gamma_n>0.1$ as a function of $N$. It is clear that
$N_{\mu} \sim N^{d/2}$ with $d=\ln{(2)} / \ln{(3)}$ the fractal dimension of the
repeller as predicted by the FWL.
%

\begin{figure}
  \includegraphics[width=8cm]{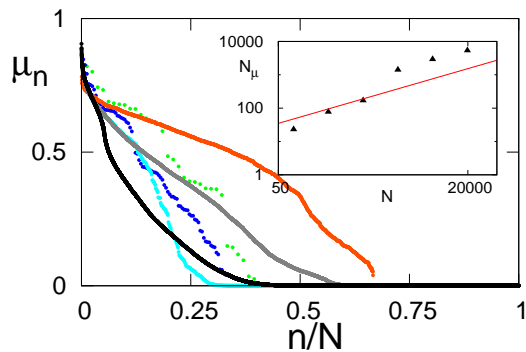}
 \caption{(color online)  Ordered decay factor $\mu_n$ of the open
  map $\tilde{B}_4^s$ as a function of $n/N$. 
  The Hilbert space dimension is $N=3^l$ with $l=4,5,6,7,8$ and $9$ (green, blue, turquoise, 
red, gray, and black symbols, respectively). 
  Inset: Fraction of states $N_{\mu}$ with decay rate $\Gamma>0.1$ as a function of 
Hilbert space dimension. The solid
red line corresponds to the prediction of the FWL ($N_\mu \sim
N^{\ln{(2)}/(2 \ln{(3)})}$).}
 \label{fig:2}
\end{figure}

In Fig. \ref{fig:2} we show the same statistics, but for the eigenvalues
of the open map $\tilde{B}_4^s$, i.e. the 4th member of the shift family. 
It is worth mentioning that we have also studied 
the other members of this family obtaining similar qualitative results as those 
shown in Fig. \ref{fig:2}. We clearly see that the ordered decay
factor for this map using different Hilbert space dimensions roughly behaves as in the case 
of the usual open tribaker map, i.e. it follows the FWL. 

\begin{figure}
  \includegraphics[width=8cm]{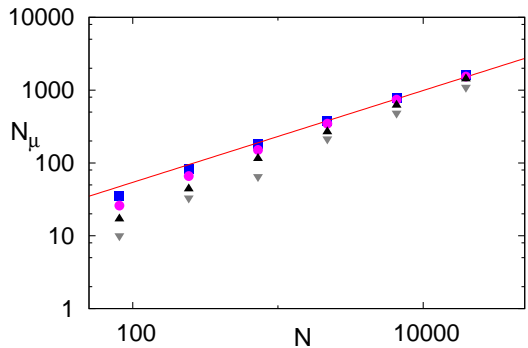}
 \caption{(color online) Fraction of states $N_{\mu}$ with decay rate $\Gamma>0.1$ 
  as a function of the Hilbert space dimension for the open maps $\tilde{B}_k^i$ 
  with $k=1,2,3,4$ (boxes, circles, up and down triangles, respectively).}
 \label{fig:3}
\end{figure}

Finally, we consider the statistics of the eigenvalues in the case of the intersection family. 
In Fig. \ref{fig:3} the  number of resonances
with $\Gamma>0.1$ as a function of $N$ for the open maps
$\tilde{B}_k^i$ with $k=1,2,3,4$ is shown.
We can see that again the number of resonances as a function of $N$ seems to follow the prediction of 
the FWL.

In the following Section we go deeper into the study of the classical support of 
open quantum maps.
\section{Transient effects and the theory of short periodic orbits for open quantum maps}
 \label{sec:shPOth}

 The recently developed theory of short POs for open quantum maps \cite{art1,art2} 
(which finds its roots in the corresponding theory for closed systems \cite{Vergini}) 
provides the ideal tool to test up to which point the repeller, more specifically the 
POs living in it are 
the actual classical support of quantum resonances. The essential ingredient of this 
approach is the open scar function associated to each one of these trajectories. 

Taking $\gamma$ as a periodic orbit of fundamental period $L$ that belongs to an open map, 
we define coherent states $|q_j,p_j\rangle$ associated to each point of the orbit 
(it has a total of $L$ points, all in the repeller) and construct the linear combination 
\begin{equation} |\phi_\gamma^m\rangle=\frac{1}{\sqrt{L}}\sum_{j=0}^{L-1}
\exp\{-2\pi i(j A^m_\gamma-N\theta_j)\}|q_j,p_j\rangle.\end{equation} In this expression 
$m\in\{0,\ldots,L-1\}$ and $\theta_j=\sum_{l=0}^j S_l$, where $S_l$ is the action acquired 
by the $l$th coherent state in one step of the map. The total action is $\theta_L\equiv S_\gamma$
and $A^m_\gamma=(NS_\gamma+m)/L$. The right and left scar functions for the periodic 
orbit are defined through the propagation of these linear combinations under the open map 
$\widetilde{U}$ (up to approximately the system's Ehrenfest time $\tau$). 
\begin{equation}\label{prop}
|\psi^R_{\gamma,m}\rangle=\frac{1}{\mathcal{N}_\gamma^R}\sum_{t=0}^{\tau}
\widetilde{U}^te^{-2\pi iA^m_\gamma t}\cos\left(\frac{\pi
t}{2\tau}\right)|\phi_\gamma^m\rangle,\end{equation} and \begin{equation}
\langle\psi^L_{\gamma,m}|=\frac{1}{\mathcal{N}_\gamma^L}\sum_{t=0}^{\tau}
\langle\phi_\gamma^m|\widetilde{U}^te^{-2\pi iA^m_\gamma t}\cos\left(\frac{\pi
t}{2\tau}\right).\end{equation} Normalization ($\mathcal{N}_{\gamma}^{R,L}$) 
is chosen in such a way that $\langle \psi_{\gamma,m}^R|\psi^R_{\gamma,m}\rangle=
\langle \psi_{\gamma,m}^L|\psi^L_{\gamma,m}\rangle$ and
$\langle \psi_{\gamma,m}^L|\psi^R_{\gamma,m}\rangle=1$.
We select a number of short POs that approximately covers the repeller, 
whose number scales following the FWL prediction \cite{art2}. 
Then, we construct an appropriate basis in which we can write open evolution operators 
associated to open maps. After solving a generalized eigenvalue problem we obtain a 
classically motivated approximation to the system's long-lived resonances \cite{art1}. 

The short periodic orbit theory is strongly relying on the repeller as the 
source of fundamental information in order to describe the long-lived 
portion of the spectra of open maps. 
In fact, it takes into account the details of the open dynamics on the repeller, 
going beyond the consideration of just a measure such as, 
for instance, the fractal dimension which governs the FWL. 
But now the question arises: Is this enough?, 
or are more elements needed?

\begin{figure}
 \includegraphics[width=8cm]{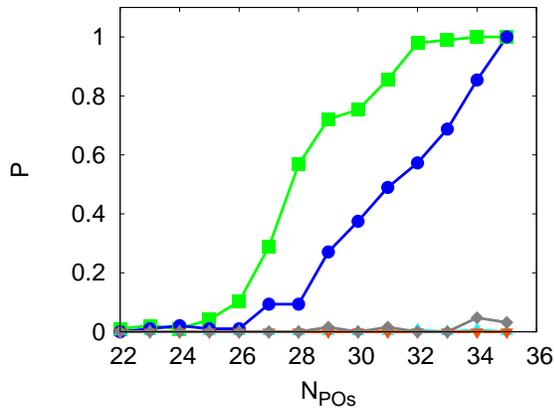}
 \caption{(color online) Performance $P$ of the short POs approach for the case of the shift family: 
fraction of long-lived resonances found as a function of the number of 
POs used in the calculations $N_{POs}$. 
Hilbert space dimension is $N=3^l$ with $l=5$, all possible $k$ are shown. (Color) gray 
lines with squares correspond to $k=1$, circles to $k=2$, triangles up to $k=3$, triangles down 
to $k=4$, and diamonds to $k=5$.}
 \label{fig:Perf_shift_l5}
\end{figure}

\begin{figure}
\includegraphics[width=8cm]{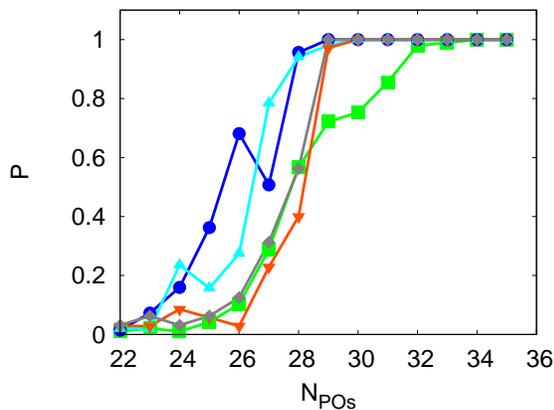}
 \caption{(color online) Performance $P$ of the short POs approach for the case of the intersection family: 
fraction of long-lived resonances found as a function of the number of 
POs used in the calculations $N_{POs}$. Hilbert space dimension is $N=3^l$ with $l=5$, all possible $k$ are 
shown. We use the same colors and symbols as in Fig. \ref{fig:Perf_shift_l5}.}
 \label{fig:Perf_intersection_l5}
\end{figure}

\begin{figure}
\includegraphics[width=8cm]{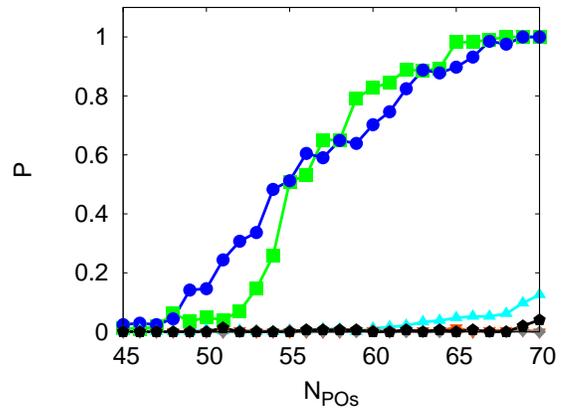}
 \caption{(color online) Performance $P$ of the short POs approach in the case of the shift family: 
fraction of long-lived resonances found as a function of the number of 
POs used in the calculations $N_{POs}$. Hilbert space dimension is $N=3^l$ with $l=6$, all possible $k$ are 
shown. (Color) gray 
lines with squares correspond to $k=1$, circles to $k=2$, triangles up to $k=3$, triangles down 
to $k=4$, diamonds to $k=5$, and pentagons to $k=6$.}
 \label{fig:Perf_shift_l6}
\end{figure}

\begin{figure}
 \includegraphics[width=8cm]{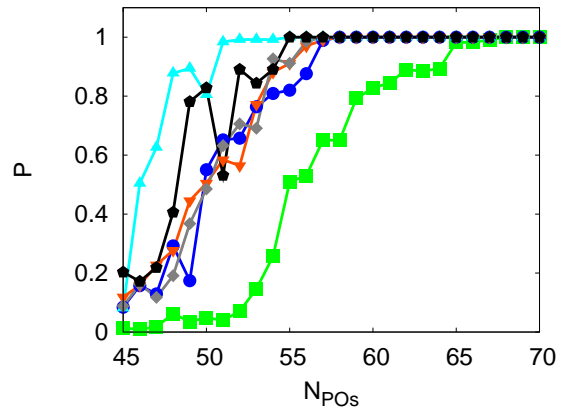}
 \caption{(color online) Performance $P$ of the short POs approach in the case of the intersection family: 
fraction of long-lived resonances found as a function of the number of 
POs used in the calculations $N_{POs}$. Hilbert space dimension is $N=3^l$ with $l=6$, all possible $k$ are 
shown. We use the same colors and symbols as in Fig. \ref{fig:Perf_shift_l6}.}
 \label{fig:Perf_intersection_l6}
\end{figure}

We define the performance $P$ of this method as the fraction of long-lived 
eigenvalues that it is able to reproduce up to an error  given by
$\epsilon=\sqrt{({\rm Re}{(z_i^{ex})}-{\rm Re}{(z_i^{PO})})^2+
({\rm Im}{(z_i^{ex})}-{\rm Im}{(z_i^{PO})})^2}$, 
where $z_i^{ex}$ and $z_i^{PO}$ are the exact eigenvalues and those given 
by the short POs theory, respectively. 
Hereafter, we take $\epsilon=0.001$ 
and consider just the eigenvalues with modulus greater than $0.01$. 
These threshold values guarantee good performance of the 
approximation and the evaluation of a meaningful number of eigenvalues 
in the whole range of maps that we have studied. 
However,  if detailed comparisons are needed,
these values should be carefully adapted to each map, this going beyond 
purposes of the present paper. 
In Figs. \ref{fig:Perf_shift_l5} and \ref{fig:Perf_intersection_l5} we show 
$P$ as a function of the number of used POs, $N_{POs}$;
here our method is applied to the shift and intersection families 
for $N=3^5$, respectively. 
At this relative low value of $N$ it is already evident that the usual 
open tribaker map that corresponds to the $k=1$ value for both families 
proves to be well reproduced by this approximation. 
In the case of the shift family the good performance is suddenly lost when 
we arrive at the $k=l/2$ threshold, i.e., from $k=3$ on, the 
method clearly fails. We notice that the resonances of the shift family 
occupy classically forbidden regions for $k \geq l/2$ \cite{art3}. 
Nevertheless, the performance is good for all the members 
of the intersection family.  We conjecture that 
this is because the probability outside the 
region roughly corresponding to the classical repeller is erased by means of the projector. 
To further confirm this conjecture we have made 
the same calculation for the $l=6$ case, which qualitatively coincides with our previous reasoning. 
The corresponding results can be seen in Figs. \ref{fig:Perf_shift_l6}
and \ref{fig:Perf_intersection_l6}.  
%

\begin{figure}
 \includegraphics[width=8cm]{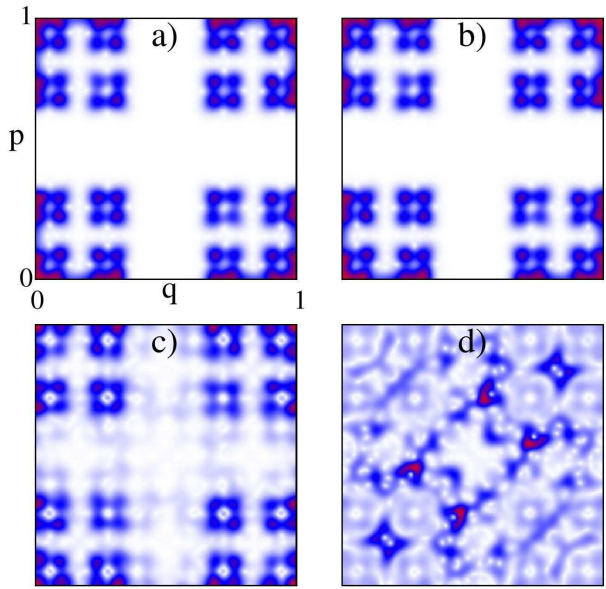}
 \caption{(color online) Phase space representation of $Q_{32}=\sum_{j=1}^{32}h_j$ for the exact 
 and short POs approach with $l=5$ and $N_{POs}=32$. 
 The map with $k=1$ (both shift and intersection families) is shown in panels a) for exact eigenstates 
 and b) for the short POs approach. For the case of shift family map with $k=3$ panel c) shows the exact eigenstates 
 and d) the short POs approach.
 The intensities of phase space representation are white for zero value and red (dark-gray) for maximum.}
 \label{fig:Qexactsemic}
\end{figure}

In order to provide further insight on the differences among members of 
these two families of maps and relate them to basic features of general open (scattering) 
quantum systems we will study the phase space distributions associated to the right 
and left eigenfunctions. For this purpose we use a recently introduced representation 
\cite{art0} that reveals the part of the quantum probability which is in the closest 
correspondence to the classical repeller. 
By defining the symmetrical operator $\hat{h}_j$ associated to the right 
$\vert R_j\rangle$ and left $\langle L_j\vert$ eigenstates
\begin{equation}\label{eq.hdef}
 \hat{h}_j=\frac{\vert R_j\rangle\langle L_j\vert}{\langle L_j\vert R_j\rangle}
\end{equation}
which is related to the eigenvalue $\lambda_j$, we construct the sum of the 
first $j$ of these projectors, ordered by decreasing modulus of the corresponding eigenvalues  
($\vert\lambda_j\vert\geqslant\vert\lambda_{j^\prime}\vert$ with $j\leq j^\prime$)
\begin{equation}
\hat{Q}_j\equiv\sum_{j^\prime=1}^j\hat{h}_{j^\prime}.
\end{equation}
The object to study consists of their phase space representation by means of 
coherent states $\vert q,p\rangle$, which is given by 
\begin{eqnarray}
 h_j(q,p)&=&\vert\langle q,p\vert \hat{h}_j\vert q,p\rangle\vert\\
 Q_j(q,p)&=& \vert\langle q,p\vert \hat{Q}_j\vert q,p\rangle\vert.
\end{eqnarray}
In Fig. \ref{fig:Qexactsemic} we show $Q_{32}$ for the exact resonances and the ones 
given by the short POs approach, obtained for $l=5$ and $N_{POs}=32$ 
(it is worth mentioning that we have observed similar results 
for $Q_{j}$ with $j$ around $32$). In the upper panels the case $k=1$ (that coincides 
for both the shift and intersection families) makes clear that the short POs theory is very 
accurate in reproducing not only the long-lived sector of the spectrum but also the component 
of the resonance eigenfunctions that live on the classical repeller. 
In fact, panels a) and b) are almost indistinguishable. 
This shows undoubtedly that the classical repeller is  
the meaningful support of long-lived states in this case.
Nevertheless, although we do not show here results for 
the intersection family, they all reproduce very 
well these components of the eigenstates,  
in agreement with the conclusions previously shown by means of the spectra. 
When the shift family is analyzed,
again a sort of phase transition behavior is observed, that 
spoils the short POs theory results in the same fashion as for the spectral values. 
This happens at $k \geq l/2$; here $l/2=3$, case which we show in 
Figs.~\ref{fig:Qexactsemic} c) and d).
%

\section{Conclusions}
 \label{sec:conclusions}

At present, it is commonly accepted that the classical information contained in the
repeller is the fundamental ingredient to describe quantum chaotic scattering
systems. 
However, a recent study \cite{art3}  has shown 
that transient features of the classical dynamics have crucial effects in the resonances 
of the quantum system.

In this paper we put this conclusion on firmer grounds. 
First, we have tested the validity of the FWL, one of the cornerstones of open chaotic
quantum systems theory in a family of quantum open tribaker maps \cite{art3}. 
We have clearly shown that the statistics of the eigenvalues follows the FWL behavior. 
In order to unveil the classical support of the resonances 
we have attempted to obtain them by means of the short POs approach. 
This theory has been recently linked to the FWL in Ref. \cite{art2}, 
and it is based on the short POs that live in the repeller. 
We have defined a measure for its performance, showing that it fails to describe 
the quantum resonances associated to $k \geq l/2$ for the shift family of maps. 
In contrast, the intersection family, whose opening forces the probability to remain 
roughly inside what can be called the classical repeller's area, 
is very well reproduced.
 
Summarizing, our results in open quantum maps have shown the need 
to include other ingredients,  
either classical and/or quantum, additionally to the repeller, 
in order to satisfactorily describe the quantum mechanics 
of scattering systems.
Further investigations are needed in order to find them and extend the 
short POs theory for open quantum maps.

\section{Acknowledgments}
This work was supported by the MICINN-Spain under contract MTM2009--14621,
CEAL, and also CONICET, UBACyT and ANPCyT (from Argentina).
%

%

\begin{thebibliography}
\eprint{}

\bibitem{Weiss} 
U. Weiss, 
{\em Quantum Dissipative Systems} (World Scientific, Singapore, 2008).

\bibitem{Nonnenmacher} 
S. Nonnenmacher, 
arXiv:1105.2457 (2011).

\bibitem{conjecture} 
W. Lu, S. Sridhar and M. Zworski,
Phys. Rev. Lett. {\bf 91}, 154101 (2003).

\bibitem{Sjostrand} 
J. Sj\:ostrand, 
Duke Math.J. {\bf 60}, 1 (1990);
J. Sj¨ostrand and M. Zworski, 
Duke Math. J. {\bf 137}, 381 (2007).

\bibitem{Hamiltonians} 
J.A. Ramilowski, S.D. Prado, F. Borondo and D. Farrelly, 
Phys. Rev. E {\bf 80}, 055201(R) (2009);
A. Ebersp\:acher, J. Main and G. Wunner, 
Phys. Rev. E {\bf 82}, 046201 (2010).

\bibitem{qmaps} 
H. Schomerus and J. Tworzydlo, 
Phys. Rev. Lett. {\bf 93}, 154102 (2004);
S. Nonnenmacher and M. Rubin, 
Nonlinearity {\bf 20}, 1387 (2007); 
D. L. Shepelyansky,
Phys. Rev. E {\bf 77}, 015202(R) (2008).

\bibitem{art1} 
M. Novaes, J.M. Pedrosa, D. Wisniacki, G.G. Carlo, and J.P. Keating, 
Phys. Rev. E {\bf 80}, 035202(R) 2009.

\bibitem{art2} 
J.M. Pedrosa, D. Wisniacki, G.G. Carlo, and M. Novaes, 
Phys. Rev. E {\bf 85}, 036203 (2012).

\bibitem{art3} 
L. Ermann, G.G. Carlo, J.M. Pedrosa, and M. Saraceno, 
Phys. Rev. E {\bf 85}, 066204 (2012).

\bibitem{bakerbvs}
N.L. Balazs and A. Voros, 
Ann. Phys. {\bf 190}, 1 (1989);
M. Saraceno, 
Ann. Phys. {\bf 199}, 37 (1990).  

\bibitem{ermannessential}
L. Ermann and M. Saraceno, 
Phys. Rev. E {\bf 74}, 046205 (2006).

\bibitem{ChaoticScattering}
P. Gaspard, Chaos, Scattering and Statistical Mechanics,
Cambridge Univ. Press, Cambridge (1998).

\bibitem{ShepelyanskyPhD99}
G. Casati, G. Maspero, and D.L. Shepelyansky,
Physica D {\bf 131}, 311 (1999).

\bibitem{KeatingPrl06}
J.P. Keating, M. Novaes, S.D. Prado, and M. Sieber,
 Phys. Rev. Lett. {\bf 97}, 150406 (2006).

\bibitem{Non1}
S. Nonnenmacher and M. Zworski,
J. Phys. A {\bf 38}, 10683 (2005).

\bibitem{Wiersig} 
J. Wiersig, 
Phys. Rev. Lett. {\bf 97}, 253901 (2006).

\bibitem{carlo} 
D.A. Wisniacki and G.G. Carlo, 
Phys. Rev. E {\bf 77}, 045201(R) (2008).

\bibitem{Microlasers}
W. Fang, Phys. Rev. A {\bf 72}, 023815 (2005); J.U. N\"ockel and
D.A. Stone, Nature (London) {\bf 385}, 45 (1997); T. Harayama,
P. Davis and K.S. Ikeda, Phys. Rev. Lett. {\bf 90}, 063901 (2003);
J. Wiersig and M. Hentschel, Phys. Rev. A {\bf 73}, 031802(R) (2006);
J. Wiersig and M. Hentschel, Phys. Rev. Lett. {\bf 100}, 033901 (2008).

\bibitem{Vergini} E. G. Vergini  J. Phys. A: Math. Gen. {\bf 33} 4709 (2000);
E. G. Vergini and G. G. Carlo, J. Phys. A: Math. Gen. {\bf 33} 4717 (2000);
E. G. Vergini, D. Schneider and A. F. Rivas, J. Phys. A: Math. Theor. {\bf 41} 405102 (2008); 
L. Ermann and M. Saraceno, Phys. Rev. E {\bf 78}, 036221 (2008).

\bibitem{art0} 
L. Ermann, G.G. Carlo, and M. Saraceno, 
Phys. Rev. Lett. {\bf 103}, 054102 (2009).

%
\end{thebibliography}
\end{document}